\documentclass{article}
\usepackage{amsmath, amssymb, amsfonts}
\title{COMMENTS ON ``THERMODYNAMICS AND/OF HORIZONS : A COMPARISON \\ OF SCHWARZSCHILD, RINDLER AND DE SITTER SPACETIMES``\\ (ArXiv : gr-qc/0202078), by T. Padmanabhan} 
\author{Hristu Culetu\\ Ovidius University, Dept.of Physics, B-dul Mamaia 124, \\ 8700 Constanta, Romania \\ email : hculetu@yahoo.com}
\begin{document}
\numberwithin{equation}{section}
\pagenumbering{arabic}
\maketitle
\newcommand{\fv}{\boldsymbol{f}}
\newcommand{\tv}{\boldsymbol{t}}
\newcommand{\gv}{\boldsymbol{g}}
\newcommand{\OV}{\boldsymbol{O}}
\newcommand{\wv}{\boldsymbol{w}}
\newcommand{\WV}{\boldsymbol{W}}
\newcommand{\NV}{\boldsymbol{N}}
\newcommand{\hv}{\boldsymbol{h}}
\newcommand{\yv}{\boldsymbol{y}}
\newcommand{\RE}{\textrm{Re}}
\newcommand{\IM}{\textrm{Im}}
\newcommand{\rot}{\textrm{rot}}
\newcommand{\dv}{\boldsymbol{d}}
\newcommand{\grad}{\textrm{grad}}
\newcommand{\Tr}{\textrm{Tr}}
\newcommand{\ua}{\uparrow}
\newcommand{\da}{\downarrow}
\newcommand{\ct}{\textrm{const}}
\newcommand{\xv}{\boldsymbol{x}}
\newcommand{\mv}{\boldsymbol{m}}
\newcommand{\rv}{\boldsymbol{r}}
\newcommand{\kv}{\boldsymbol{k}}
\newcommand{\VE}{\boldsymbol{V}}
\newcommand{\sv}{\boldsymbol{s}}
\newcommand{\RV}{\boldsymbol{R}}
\newcommand{\pv}{\boldsymbol{p}}
\newcommand{\PV}{\boldsymbol{P}}
\newcommand{\EV}{\boldsymbol{E}}
\newcommand{\DV}{\boldsymbol{D}}
\newcommand{\BV}{\boldsymbol{B}}
\newcommand{\HV}{\boldsymbol{H}}
\newcommand{\MV}{\boldsymbol{M}}
\newcommand{\be}{\begin{equation}}
\newcommand{\ee}{\end{equation}}
\newcommand{\ba}{\begin{eqnarray}}
\newcommand{\ea}{\end{eqnarray}}
\newcommand{\bq}{\begin{eqnarray*}}
\newcommand{\eq}{\end{eqnarray*}}
\newcommand{\pa}{\partial}
\newcommand{\f}{\frac}
\newcommand{\FV}{\boldsymbol{F}}
\newcommand{\ve}{\boldsymbol{v}}
\newcommand{\AV}{\boldsymbol{A}}
\newcommand{\jv}{\boldsymbol{j}}
\newcommand{\LV}{\boldsymbol{L}}
\newcommand{\SV}{\boldsymbol{S}}
\newcommand{\av}{\boldsymbol{a}}
\newcommand{\qv}{\boldsymbol{q}}
\newcommand{\QV}{\boldsymbol{Q}}
\newcommand{\ev}{\boldsymbol{e}}
\newcommand{\uv}{\boldsymbol{u}}
\newcommand{\KV}{\boldsymbol{K}}
\newcommand{\ro}{\boldsymbol{\rho}}
\newcommand{\si}{\boldsymbol{\sigma}}
\newcommand{\thv}{\boldsymbol{\theta}}
\newcommand{\bv}{\boldsymbol{b}}
\newcommand{\JV}{\boldsymbol{J}}
\newcommand{\nv}{\boldsymbol{n}}
\newcommand{\lv}{\boldsymbol{l}}
\newcommand{\om}{\boldsymbol{\omega}}
\newcommand{\Om}{\boldsymbol{\Omega}}
\newcommand{\Piv}{\boldsymbol{\Pi}}
\newcommand{\UV}{\boldsymbol{U}}
\newcommand{\iv}{\boldsymbol{i}}
\newcommand{\nuv}{\boldsymbol{\nu}}
\newcommand{\muv}{\boldsymbol{\mu}}
\newcommand{\lm}{\boldsymbol{\lambda}}
\newcommand{\Lm}{\boldsymbol{\Lambda}}
\newcommand{\opsi}{\overline{\psi}}
\renewcommand{\tan}{\textrm{tg}}
\renewcommand{\cot}{\textrm{ctg}}
\renewcommand{\sinh}{\textrm{sh}}
\renewcommand{\cosh}{\textrm{ch}}
\renewcommand{\tanh}{\textrm{th}}
\renewcommand{\coth}{\textrm{cth}}

~~In the paper \cite{TP}, T.Padmanabhan has used the class of metrics
\begin{equation}
ds^{2} = -f(r) dt^{2}+f^{-1}(r) dr^{2}+dL_{\bot}^{2}
\label{1}
\end{equation}
as a canonical ensemble at constant temperature $T=1/\beta$, with $dL_{\bot}^{2}$- the transverse 2-dimensional metric. \\
~~He calculated the partition function 
\begin{equation}
Z(\beta)\propto \exp(S-\beta E)
\label{2}
\end{equation}
where S and E are the entropy and energy of the system, respectively.\\ The function f(r) is vanishing at some surface $r=a$ (event horizon) and $f'(a)$ is finite. In this case $T= |f'(a)|/4 \pi$. Padmanabhan computed S and E using the expression of $Z(\beta)$ in terms of the (Euclidean) gravitational action (the metric is static) and found that 
\begin{equation}
S = \frac{1}{4} 4 \pi a^{2}= \frac{A_{hor}}{4} ; ~~~~E = \frac{a}{2}
\label{3}
\end{equation}
~For a spacetime with planar symmetry (for instance, the Rindler geometry), the author established that $S = (1/4) A_{\bot}$ while the energy is vanishing. \\
~~We try in this note to prove that the expression $E = a/2$ is valid for the Rindler spacetime, too.\\
~F.Alexander and U.Gerlach \cite{AG} showed that the electric field of a uniformly accelerated charge $e$ induces on the event horizon a surface charge density such that the total charge on that surface is $-e$. In other words, the event horizon behaves as a conductive surface. Moreover, an attractive force between the point charge and the horizon appears, its expression proving to be equal to the radiation reaction force from the r.h.s. of Dirac's equation.\\
 By analogy with the electromagnetic case, we suggest to consider that the Rindler horizon contains an (observer dependent) surface density $\sigma$ given by 
\begin{equation}
g = 4 \pi G \sigma.
\label{4}
\end{equation}
where g is the proper acceleration of the hyperbolic observer.\\
From now on we take $G = \hbar = c = k_{B} = 1$.\\
Among others, Padmanabhan \cite{TP1} (see also \cite{MP}) has observed that the relation (4) ``makes physical sense because the accelerated observer will attribute such a surface energy density as the source of apparent gravitational acceleration``. It is, however, in contradiction with his $E = 0$ result obtained in the paper \cite {TP} for the Rindler spacetime.\\
To show that $E = a/2$ is true also for the Rindler horizon, we look for an analogy with Schwarzschild's geometry
\begin{equation}
ds^{2} = -(1-\frac{2m}{r}) dt^{2}+\frac{dr^{2}}{1-\frac{2m}{r}}+ dL_{\bot}^{2}
\label{5}
\end{equation}
~~It is well known \cite{TP2} that near the horizon $r = 2m$ of the black hole, (5) becomes approximately the Rindler spacetime
\begin{equation}
ds^{2} = -\frac{r-2m}{2m} dt^{2}+\frac{2m}{r-2m} dr^{2}+ dL_{\bot}^{2}
\label{6}
\end{equation}
~By means of the coordinate transformation
\begin{equation}
\sqrt{r-2m} = \frac{\rho}{2 \sqrt{2m}}
\label{7}
\end{equation}
eq. (6) becomes
\begin{equation}
ds^{2} = -(\frac{1}{4m})^{2} \rho^{2} dt^{2}+d \rho^{2}+dL_{\bot}^{2}
\label{8}
\end{equation}
The proper acceleration is $g = 1/4m$, i.e. exactly the surface gravity of the black hole.\\
~~Let us consider now a slightly different transformation. Instead of (7) we could introduce
\begin{equation}
\sqrt{r-2m} = \frac{\rho-4m}{2 \sqrt{2m}}
\label{9}
\end{equation}
Namely, we take $\rho = 1/g$ for $r = 2m$. In this case eq.(8) looks like 
\begin{equation}
ds^{2} = -(g \rho-1)^{2} dt^{2}+d\rho^{2}+dL_{\bot}^{2} 
\label{10}
\end{equation}  
~In other words, the horizon is translated from $\rho = 0$ to $\rho = 1/g$, with $\rho \geq 1/g$. \\
~~It is clear that the surface gravity $\kappa = 1/4m$ of the black hole plays a similar role with the proper acceleration g of the uniformly accelerated observer. Keeping in mind that the energy of the black hole is given by $E = m = a/2 = 1/4 \kappa$, we have therefore for the Rindler horizon
\begin{equation}
E(g) = \frac{1}{4g} = \frac{a}{2}
\label{11}
\end{equation}
(We have, of course, tacitly assumed that all the energy of the black hole is located at the horizon $r = a$. A similar situation is faced when the Vilenkin - Ipser - Sikivie (VIS) \cite{IS} \cite{AV} domain wall is studied. As Chamblin and Eardley \cite{CE} have noticed, ``we think of a VIS spacetime as an inflating universe where all of the vacuum energy has been concentrated on the sheet of the domain wall``). \\
~With all fundamental constants, eq. (11) appears as $E = (c^{4}/4G)~(c^{2}/g)$.\\
~ The surface energy density on the horizon looks now as 
\begin{equation}
\sigma = \frac{1}{4g} \frac{1}{4 \pi (4m^{2})} = \frac{g}{4 \pi}
\label{12}
\end{equation}
whence $g = 4 \pi \sigma$, as expected (for instance, from Gauss' theorem \cite {MP})\\
~The fact that Rindler's horizon might contain certain energy is also stressed by T.Padmanabhan in \cite{TP3} where he proved that the scalar curvature is nonvanishing and is concentrated at the horizon when we treat the Rindler geometry as a limiting case of a family of metrics without horizons (see also \cite{HC1}).\\
~~Having established that the energy of the ``horizon membrane`` is given by (11), we pass now to the thermodynamical parameters. The temperature is, of course, the Unruh temperature $T_{U} = g/2 \pi$. For the entropy of the Rindler horizon we use the ``holographic`` expression \cite{HC2} $S = A/4$, ``A`` being the horizon area. One obtains
\begin{equation}
S(g) = 4 \pi m^{2} = \frac {\pi}{4g^{2}}
\label{13}
\end{equation}
~It is easy to check that the thermodynamic relation $dE = T dS$ is obeyed.\\
~Even though we started with an analogy between the Schwarzschild and Rindler spacetimes near the black hole horizon, we believe the Rindler horizon contains a surface energy given by eq. (12) and generated by the agent who accelerates the test particle. We also stress that eq. (11) preserves its form even for the planar horizon of an uniformly accelerated observer, in Cartesian coordinates. This might be justified basing on the fact that, for an observer located very near the black hole horizon, it appears as a flat surface (the linear dimension of the observer is considered to be much less than the Schwarzschild radius).\\
~In our coordinates (10), it looks as a ``bubble`` (domain wall) located at a distance $\rho = 1/g$. When the expression (12) for $\sigma$ is written as 
\begin{equation}
\sigma = \frac {T}{2 l_{P}^{2}}
\label{14}
\end{equation}
where $l_{P}$ is the Planck length, we observe that in every ``surface element`` $l_{P}^{2}$ we find a normal mode average energy of $T/2$. \\
~It is worth to notice a relation between the classical expression (11) for the Rindler energy and the quantum one \cite{HC3}
\begin{equation}
E_{quan} = \frac{\hbar c}{2 \frac{c^{2}}{g}}.
\label{15}
\end{equation}
Since $E_{class} \propto 1/g$ and $E_{quan} \propto g$, we have 
\begin{equation}
 E_{quan}~ E_{class} \approx \epsilon_{P}^{2}.
\label{16}
\end{equation}
~In other words, the Planck energy $\epsilon_{P}$ is of the order of the geometrical mean of the two, playing the role of a ``boundary`` between the two domains of energies.\\ 
~~To summarize, we showed in this note that Padmanabhan's expression $E = a/2$ for the Schwarzschild and de Sitter energy is valid also for the Rindler horizon, even in Cartesian coordinates. In addition, the corresponding entropy $S(g)$ is finite and the thermodynamic relation $dE = T dS$ is fulfilled.\\
Acknowledgements\\
~I am grateful to Prof. T.Padmanabhan for useful discussions and for suggesting me to write this note.

\end{document}